\documentclass[useAMS,usenatbib]{mn2e}
\usepackage{myaasmacros}
\input{epsf}

\def\ltsima{$\; \buildrel < \over \sim \;$}
\def\simlt{\lower.5ex\hbox{\ltsima}}   
\def\gtsima{$\; \buildrel > \over \sim \;$}
\def\simgt{\lower.5ex\hbox{\gtsima}}

\title[Mond and the Sagittarius stream]
{Tidal streams in a MOND potential: constraints from Sagittarius}

\author[Read \& Moore]{J. I. Read $^1$\thanks{Email: jir22@ast.cam.ac.uk} 
\& Ben Moore $^2$ 
\\ $^1$Institute of Astronomy, Cambridge University, Madingley Road, 
Cambridge, CB3 OHA, England
\\ $^2$Institute of Theoretical Physics, 
University of Z\"urich,  
Wintherurestrasse 190, 8057 Zurich, Switzerland}

\begin{document}

\maketitle

\begin{abstract}
We compare orbits in a thin axisymmetric disc potential in MOND to
those in a thin disc plus near-spherical dark matter halo predicted by
a $\Lambda$CDM cosmology. Remarkably, the amount of orbital
precession in MOND is nearly identical to that which occurs in 
a mildly oblate CDM Galactic halo (potential flattening q=0.9),
consistent with recent 
constraints from the Sagittarius stream. Since very flattened mass
distributions in MOND produce rounder potentials than in
standard Newtonian mechanics, we show that it will be very difficult
to use the tidal debris from streams to distinguish between a MOND
galaxy and a standard CDM galaxy with a mildly oblate halo.

If a galaxy can be found with either a prolate halo, or one which is
more oblate than $q \sim 0.9$ this would rule out MOND as a viable
theory. Improved data from the leading arm of the Sagittarius dwarf -
which samples the Galactic potential at large radii - could rule out
MOND if the orbital pole precession can be determined to an accuracy
of the order of $\pm 1^o$.

\end{abstract}


\section{Introduction}\label{sec:introduction}

Ever since Zwicky's seminal work in the 1930's it has been known that
there is a disparity between the mass of galaxies as measured
dynamically and the mass inferred from the visible light.
The standard explanation for this missing matter is to invoke one or
many weakly interacting massive particles which form early in the
universe and to first order interact only via gravity (see
e.g. \citet{1998NewAR..42..245B}). This is known as cold dark matter
or CDM theory. However, since none of the well
motivated candidate particles has yet been detected, it is important
to also consider alternative theories as an explanation for the missing
matter.

One such alternative theory, is a modification of standard Newtonian
gravity (MOND) for accelerations below some characteristic scale, $a_0
\sim 1.2 \times 10^{-10}$ms$^{-2}$ (\citet{1983ApJ...270..365M} and \citet{2004ApJ...611...26M}). MOND was
first suggested by Milgrom in 1983 as a modified inertia theory, but
since then has been expanded into a self consistent Lagrangian field
theory \citep{1984ApJ...286....7B} and, more recently, has been placed
on a firm footing within the context of general relativity
\citep{2004PhRvD..70h3509B}. This last point is of particular interest
since \citet{2004PhRvD..70h3509B} has managed to address many of the
conceptual problems that have plagued MOND over the past two decades
and shown that the theory can be consistent with gravitational lensing
and other general-relativistic phenomena.

In this paper we compare the potential of the Milky Way as predicted
by MOND and CDM models. In the former the potential arises from a 
flattened disk of baryons whereas the latter potential is primarily
from an extended spheroidal distribution of dark matter.
A perfectly spherical potential has orbits which are confined to lie
on planes \citep{1987gady.book.....B}. By contrast, orbits in
axisymmetric potentials generally show precession of their orbital planes (this
will be true for all orbits which are not exactly planar or exactly polar).

\citet{2001ApJ...551..294I} and more recently
\citet{2003ApJ...599.1082M}, \citet{2005ApJ...619..800J} and 
\citet{2005ApJ...619..807L} have studied the tidal debris from the
Sagittarius dwarf galaxy and calculated likely orbits for the
galaxy and its stellar debris. 
They find that the precession of the orbital plane is small
($\sim 10^o$) and is consistent with a galactic halo potential which
is only mildly oblate ($q = 0.9-0.95$) (although see also
\citet{2004ApJ...610L..97H} and section \ref{sec:discussion} in this
paper). In MOND, where all of the gravity comes
only from the disc, the potential may be much flatter leading to far more
precession than is observed. In this way, tidal debris from
infalling satellites such as the Sagittarius dwarf could provide
strong constraints on any altered gravity theory which supposes that
all gravity is produced only by the visible light. 

This paper is organised as follows: In section \ref{sec:mp} we discuss
the Galactic MOND potential.
In section \ref{sec:initial} we
outline the initial conditions and orbit solver. We use four models:
the CDM model which has a spherical dark matter halo, the f095CDM and
f09CDM models
which have slightly oblate halos ($q=0.95$ and $q=0.9$) and the MOND
model, where 
all of the gravitational potential comes from the disc. 
In section \ref{sec:results} we
present our results for two different orbits: the first is motivated
by the orbit of the Sagittarius dwarf and the second is a
small-pericentre orbit chosen to sample a wide range of the Galactic
potential. In section
\ref{sec:discussion} we briefly discuss the significance of these
results and relate our work to previously published studies. Finally,
in section \ref{sec:conclusion} we present our conclusions.

\section{The MOND potential}\label{sec:mp}
 
The MOND field equations lead to a modified, non-linear, version of
Poisson's equation given by \citep{1984ApJ...286....7B}:

\begin{equation}
\underline{\nabla} \cdot[\mu(|\underline{\nabla}(\Phi)|/a_0)\underline{\nabla}\Phi]=4\pi G \rho
\label{eqn:mondpoisson}
\end{equation}
Where $\rho$ is the density, $\Phi$ is the scalar field for MONDian
gravity and $a_0$ is the acceleration scale below which gravity
deviates from standard Newtonian behaviour. The unknown function
$\mu(|\nabla(\Phi)|/a_0)$ parameterises the change from Newtonian to
MONDian gravity and is usually given phenomenologically by
$\mu(x)=x(1+x^2)^{-1/2}$ \citep{1984ApJ...286....7B}.

Equation \ref{eqn:mondpoisson} is in general extremely difficult to
solve, not least because it is trivial to show that substituting $\Phi
\rightarrow \Phi_1+\Phi_2$ does not give $\rho \rightarrow
\rho_1+\rho_2$. This means that solutions cannot be superposed as in
normal Newtonian mechanics. Every mass configuration will have its
own unique potential which should be determined by (numerically)
inverting equation \ref{eqn:mondpoisson}. Thus, while some authors
(see e.g. \citet{2004MNRAS.347.1055K}) have made valiant efforts to
adapt N-body integrators to work in MOND, these can only be, at best,
an approximation.

However, equation \ref{eqn:mondpoisson} may be solved in extremely
special cases. Following \citet{1995MNRAS.276..453B}, notice that we
may write the MONDian gravitational field ($\underline{g}=\underline{\nabla}\Phi$) as
the sum of the Newtonian gravitation field
($\underline{g_N}=\underline{\nabla}\Phi_N$) and a curl field: 

\begin{equation}
\mu(|\underline{g}|/a_0) \underline{g} = \underline{g_N}+\underline{\nabla}\times\underline{h}
\label{eqn:curlfield}
\end{equation}

The curl field will trivially vanish for planar, spherical or
cylindrical symmetry giving (in exact agreement with the modified
inertia interpretation of MOND \citep{1983ApJ...270..365M}):

\begin{equation}
\mu(|\underline{g}|/a_0) \underline{g} = \underline{g_N}
\label{eqn:mondinertia}
\end{equation}

Which, substituting for $\mu(x)=x(1+x^2)^{-1/2}$ as above and inverting
gives:

\begin{equation}
\underline{g} = \underline{g_N}\frac{\left(1+\sqrt{1+4a_0^2/|\underline{g_N}|^2}\right)^{1/2}}{\sqrt{2}}
\label{eqn:mondequation}
\end{equation}

Equation \ref{eqn:mondequation} is much more tractable since
$\underline{g_N}$ may be calculated from the Newtonian potential as usual and
then simply modified to give the correct MONDian acceleration at a
given point in the field. 

Our galaxy is clearly neither planar, spherical nor cylindrical and so
the applicability of equation \ref{eqn:mondequation} may rightly be
questioned. However, \citet{1995MNRAS.276..453B} demonstrated that
equation \ref{eqn:mondequation} may be used {\it exactly} for
infinitesimally thin Kuzmin discs with {\it Newtonian} potential
given by \citep{1987gady.book.....B}:

\begin{equation}
\Phi_N(R,z) = \frac{-GM}{\sqrt{R^2+(a+|z|)^2}}
\label{eqn:kuzmin}
\end{equation}
Where $G$ is the gravitational constant, $a$ is the disc scale length
and $M$ is the mass of the disc. The reason that the Kuzmin potential
can be used exactly is because it is an extremely special potential
for which $|\nabla\Phi_N|=f(\Phi_N)$\footnote{For the Kuzmin potential,
  $|\nabla\Phi_N|=\Phi_N^2/GM$.}. Notice that in MOND, the force field
is still the gradient of a scalar potential and so the MONDian field must be
conservative; that is: $\underline{\nabla} \times \underline{g}=\underline{0}$. Thus a MONDian
field may be generated via equation \ref{eqn:mondinertia} from a
Newtonian field provided that the Newtonian field satisfies the
following constraint:
$\underline{\nabla}|\underline{\nabla}\Phi_N|\times\underline{\nabla}\Phi_N=\underline{0}$. This is
satisfied exactly by the Kuzmin disc.

In this paper we use the Kuzmin potential to study orbits in
axisymmetric potentials in MOND. We compare these orbits to similar
orbits in standard Newtonian mechanics (the CDM model) using the
Kuzmin potential plus a flattened spherical logarithmic potential (to
model the dark matter) given by \citep{1987gady.book.....B}:

\begin{equation}
\Phi_{L}(R,z) = \frac{1}{2}v_0^2\ln\left(R_c^2+R^2+\frac{z^2}{q^2}\right)+constant
\label{eqn:loghalo}
\end{equation}
Where $R_c$ is the scale length, $0.7 \le q \le 1$ is the halo
flattening and $v_0$ is the asymptotic value of the circular speed of
test particles at large radii in the halo.

\begin{figure} 
\begin{center}
\epsfxsize=8truecm \epsfbox{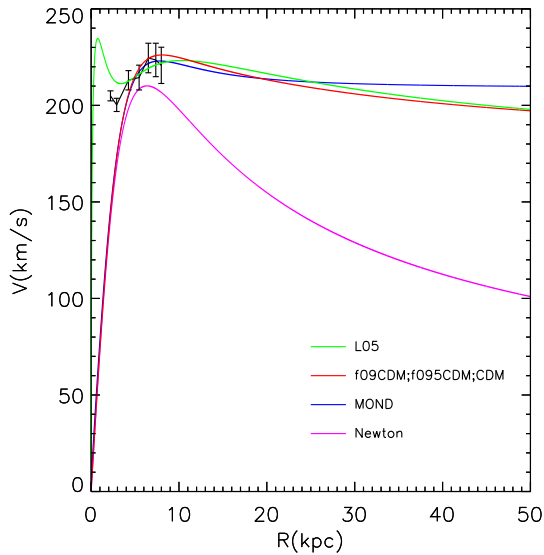}
  \caption[]
  {Rotation curves for the MOND model (blue line), the CDM models (red
    line) and the L05 model (green line). The rotation curve which the
    Kuzmin disc would give {\it without} MOND is also overplotted for
    comparison (magenta line). The black data points show the mean of
    HI measurements of the rotation curve taken from
    \citet{1984ApJS...54..513B}, \citet{1974AAS...17..251W},
    \citet{1973AAS....8....1W}, \citet{1995ApJ...448..138M} and
    \citet{1986AAS...66..373K}.}
\label{fig:rotationcurves}
\end{center}    
\end{figure}

We will also compare these to the more realistic Galactic potential
used by \citet{2005ApJ...619..800J} and
\citet{2005ApJ...619..807L}. They use a logarithmic halo (equation
\ref{eqn:loghalo}), a Miyamoto-Nagai potential for the disc
\citep{1987gady.book.....B} and Hernquist potential for the bulge
\citep{1990ApJ...356..359H}:

\begin{equation}
\Phi_{disc} = \frac{-GM}{\sqrt{R^2+(a+\sqrt{z^2+b^2})^2}}
\label{eqn:johndisc}
\end{equation}
\begin{equation}
\Phi_{bulge} = \frac{-GM_{bulge}}{r+c}
\label{eqn:johnbulge}
\end{equation}
where $a$ is the disc scale length as in equation \ref{eqn:kuzmin},
$b$ is the disc scale height, $M$ is the disc mass, $c$ is the
bulge scale length and $M_{bulge}$ is the bulge mass. Notice that for
$b \rightarrow 0$ equation \ref{eqn:johndisc} reduces to the Kuzmin
disc in equation \ref{eqn:kuzmin}.

\section{Initial conditions and orbit solving}\label{sec:initial}
The mass distribution we use in MOND is the flattened Kuzmin disc (see
equation \ref{eqn:kuzmin}). For the CDM model we use a Kuzmin disc plus a
logarithmic halo (see equation \ref{eqn:loghalo}). We present three CDM
models: one with no halo flattening (CDM), one with $q=0.95$
(f095CDM) and one with $q=0.9$ (f09CDM). We also compare these with
the best fit Milky Way potential from \citet{2005ApJ...619..800J} and
\citet{2005ApJ...619..807L} (L05). The parameters used
in all five models are given in table \ref{tab:initial} and are chosen to
match the measured rotation curve 
of the Milky Way. 

Figure \ref{fig:rotationcurves} shows the rotation
curve for the MOND model (blue line), the CDM models\footnote{All of
  the CDM models will produce the same rotation curve since the force
  from the Galaxy on the satellite {\it in the plane} of the Galaxy is
  independent of the dark matter halo flattening, $q$.} (red line) and
the L05 model (green line). The rotation curve which the Kuzmin disc
would give {\it without} MOND is also overplotted for comparison
(magenta line). The black data points show the mean of HI measurements
of the rotation 
curve taken from \citet{1984ApJS...54..513B}, \citet{1974AAS...17..251W},
\citet{1973AAS....8....1W}, \citet{1995ApJ...448..138M} and
\citet{1986AAS...66..373K}. It is important to note that
we are not attempting to form an accurate model of the Milky Way in
this paper, but instead we wish only to compare orbits in MOND and
CDM galaxies. The Kuzmin potential is not the most accurate model for
the stellar distribution of the Milky Way (see
e.g. \citet{1981ApJ...251...61C}) and in both the CDM and MOND models,
we have made no attempt to model
the stellar bulge and bar although it is 
well known that they contribute significantly to the potential of the galaxy
(\citet{1981ApJ...251...61C} and \citet{1995ApJ...445..716D}). This
can be seen in the difference between the L05 rotation curve with a bulge
component (green line) and all of the other models. Interior to
$\sim 10$kpc, all of the rotation curves deviate quite strongly from
L05. However, since most Milky Way satellites orbit well outside of 10kpc, a
potential model for the Milky Way which is accurate beyond this point
should suffice.

\begin{table*}
\begin{center}
\setlength{\arrayrulewidth}{0.5mm}
\begin{tabular}{llllllllll}
\hline
{\it Model} & {\it $M$(M$_\odot$)} & {\it $a$(kpc)} & {\it $b$(kpc)}
& {\it $v_0$(km/s)} & {\it
  $R_c$(kpc)} & {\it $q$} & {\it $M_{bulge}$(M$_\odot$)} & {\it $c$(kpc)} & {\it $a_0$(ms$^{-2}$)} \\
\hline
MOND & $1.2\times 10^{11}$ & $4.5$ & - & - & - & - & - & - & $1.2 \times 10^{-10}$\\
CDM & $1.2\times 10^{11}$ & $4.5$ & - & 175 & 13 & 1 & - & - & - \\
f095CDM & $1.2\times 10^{11}$ & $4.5$ & - & 175 & 13 & 0.95 & - & - & - \\
f09CDM & $1.2\times 10^{11}$ & $4.5$ & - & 175 & 13 & 0.9 & - & - & - \\
L05 & $1\times 10^{11}$ & $6.5$ & $0.26$ & 171 & 13 & 0.9 & $3.53\times 10^{10}$ & 0.7 & - \\
\hline
\end{tabular}
\end{center}
\label{tab:initial}
\caption[]{Initial conditions.}
\end{table*}

\begin{figure*} 
\begin{center}
\epsfxsize=18truecm \epsfbox{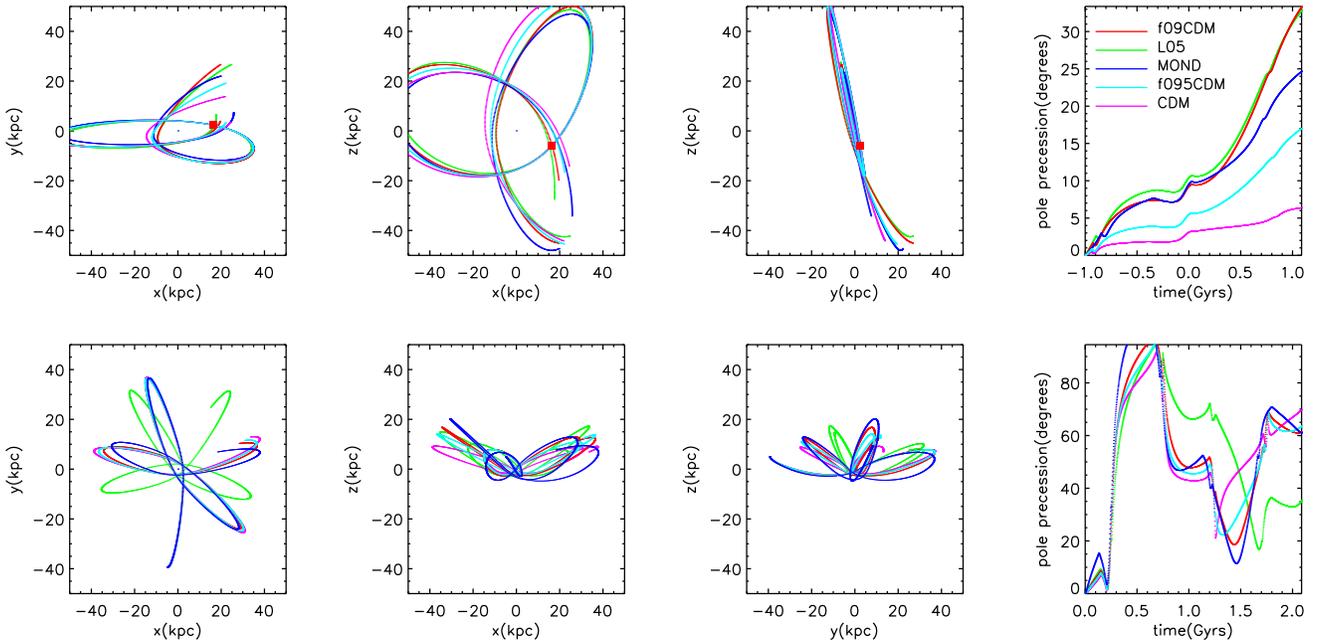}

  \caption[]
  {Orbital projections for five models as shown in the legend, 
    integrated over 2Gyrs. The galactic 
    plane (not marked) is perpendicular to the z-axis. The top panels
    are for the best-fit Sagittarius dwarf orbit which is on a
    near-polar orbit around the galaxy \citep{2005ApJ...619..807L}. The
    bottom panels are for a small-pericentre orbit which samples a
    wide range of the Galactic potential. The position of the
    Sagittarius dwarf now is marked on the top panels with a red
    square. The orbital pole precession is shown in the right-most
    panels. For the Sagittarius dwarf orbit, the time marked is
    relative to its current phase space position. The observed
    trailing arms then trace out the orbital path the dwarf took over
    the past Gyr (hence the {\it negative} time), while the leading
    arms show what its path will be over the next Gyr (hence the
    positive time).}
\label{fig:mondorbits}
\end{center}    
\end{figure*}

The equation of motion in MOND and in CDM is given by:

\begin{equation}
\underline{\ddot{x}} = -\underline{\nabla}\Phi
\label{eqn:eqmotion}
\end{equation}
Where in MOND $\underline{\nabla}\Phi=\underline{g}$ is calculated from equation
\ref{eqn:mondequation}. 

Equation \ref{eqn:eqmotion} represents a set of coupled differential
equations which we solve numerically using the fourth-order
Runge-Kutta technique \citep{1992nrca.book.....P} with a timestep of
$0.15$Myrs. Reducing the time step was found to produce converged
results, while for purely spherical potentials, the code was found to
conserve energy and angular momentum to machine accuracy (better than
1 part in $10^7$)\footnote{The orbits in the axisymmetric potentials
  used in this paper also conserved energy and the z-component of the angular
  momentum to machine accuracy - as expected for potentials with
  axisymmetry \citep{1987gady.book.....B}.}.

\section{Results}\label{sec:results}

We modelled the Sagittarius dwarf orbit by fitting to the best-fit
orbit presented in \citet{2005ApJ...619..807L} and ensuring that the
current position and velocity of the dwarf matched observational
constraints. This gives a current phase space position and velocity of
the dwarf (in Galacto-centric {\it right-handed} coordinates) of
$x=16.2,y=2.3,z=-5.9,vx=238,vy=-42,vz=222$, where numbers are quoted
in units of kpc and km/s respectively. This satellite phase space
position was then integrated backwards 1Gyr in time to match the
trailing arm of the Sagittarius dwarf and forwards 1Gyr in time to
match the leading arm.  Unlike \citet{2005ApJ...619..807L}, we did not
allow the final phase space position of the dwarf to vary within
observational constraints, but held this fixed. This is because we
wish to measure the difference in orbital precession between the
models, which is easier to do if the initial phase space coordinates
are identical.

Figure \ref{fig:mondorbits} shows orbital projections for all five
models as shown in the legend, integrated over 2Gyrs. The galactic 
    plane (not marked) is perpendicular to the z-axis. The top panels
    are for the best-fit Sagittarius dwarf orbit which is on a
    near-polar orbit around the galaxy \citep{2005ApJ...619..807L}. The
    bottom panels are for a small-pericentre orbit which samples a
    wide range of the Galactic potential. The position of the
    Sagittarius dwarf now is marked on the top panels with a red
    square. The orbital pole precession is shown in the right-most
    panels. This measures the difference in angle between the vector
    perpendicular to the satellite's orbit initially and at a given
    time. 

All of the models produced very similar orbits for the Sagittarius
dwarf and the differences are best seen in the right-most panel plots
of the orbital pole precessions. The pole precession is a useful
quantity to measure for the orbits because it is a strong function of
how flattened a potential is - this is why the differences between the
orbits in each of the models shows up so strongly in the plot of the
pole precession, whereas it is much harder to detect in the plots of
the orbital projections. Small differences in the orbits
between models may be accounted for by altering the details of the
{\it visible} component of the Milky Way potential (recall that the Kuzmin
disc used for the MOND model is only an approximation to the true
potential of the Milky Way disc and bulge); or by altering the final
phase space position of the 
Sagittarius dwarf within observational constraints as was done by
\citet{2005ApJ...619..807L}. The pole precession, however, may only
be reproduced by changing how flat the potential is. In the CDM
models, this may be achieved quite easily by using a more oblate dark
matter halo. In MOND there is less freedom to do this. While
changing the mass of the Milky Way disc can produce more or less
precession, there are strong limits from stellar population models and
from the rotation curve of the Milky Way as to how
much this can be done. If MOND produces far too much, or far too little
precession as compared with the Sagittarius stream, then we can rule
it out as a viable alternative theory to dark matter.

The CDM model (magenta line) produced
the least precession as can be expected for a near-spherical potential (recall
that this model still contains a massive disc and so we should still expect
some precession). The f09CDM and L05 models produced very similar
results. This is because, with an orbital pericentre greater than
10kpc, the Sagittarius dwarf is not sampling a region of the Milky Way
potential where the presence of the bulge is significant. The key
point is that, surprisingly, the MOND model (blue line) produced {\it
  near-identical pole precession} to both the L05 and f09CDM models -
consistent with the best fitting orbit for the Sagittarius dwarf
debris. If anything, the MOND model produced {\it too little}
precession for the leading arm of the Sagittarius dwarf debris. This
result seems surprising since in MOND all of the 
gravity is coming from the disc. We will discuss this further in
section \ref{sec:discussion}. 

There could, then, be an early indication that MOND is inconsistent
with the Sagittarius stream orbit. However, current data from the
Sagittarius dwarf is only good enough to constrain the pole precession
over the period -0.4 to 0.6Gyrs with an error of 2-3 degrees
\citep{2005ApJ...619..800J}. This places a flattening of $q=0.9-0.95$
within the 1.5$\sigma$ error bars. Yet the difference in pole
precession between $q=0.9$ and $q=0.95$ is much larger than the
difference between MOND and the f09CDM or L05 models. Over the range -0.4
to 0.6Gyrs discussed in \citet{2005ApJ...619..800J}, MOND produces
near-identical results to the f09CDM and L05 models.

There is still some lively debate about what the best fitting orbit for the
Sagittarius dwarf actually is (see section
\ref{sec:velocity}). However, it seems unlikely that it would be
possible to produce {\it more} precession in MOND than that from an
infinitesimally thin disc (the potential used in this paper). If this
is true, then it may be possible in the future to rule out MOND on the
grounds that it {\it does not produce enough} precession to match the
Sagittarius stream data.

The bottom panels show a more extreme orbit. Notice that now the orbit
for the L05 model strongly deviates from the others. This is because
this satellite orbit has a pericentre of $\sim 2$kpc and so the
satellite is now sampling the region of the Milky Way potential where
the bulge makes a difference. The rest of the models, which have no
bulge component, once again show very similar orbits. The pole
precession in MOND is again best matched by the f09CDM model, while
the f095CDM and CDM models show less precession, as is expected from
their rounder halo potentials.

The MOND potential produces orbits which are very similar to those in
a CDM halo with flattening of $q\sim 0.9$. This seems to be the case
even for orbits with unrealistically small pericentres. This suggests
that, for the Milky Way at least, it will be difficult to distinguish
between MOND and a dark matter model for the Galaxy using satellite
streams. Even if better data could be obtained for the globular
cluster orbits (which orbit closer to the Galaxy), MOND can be
expected to produce similar results to a mildly oblate dark matter halo. 

\begin{figure*} 
\begin{center}
\epsfxsize=18truecm \epsfbox{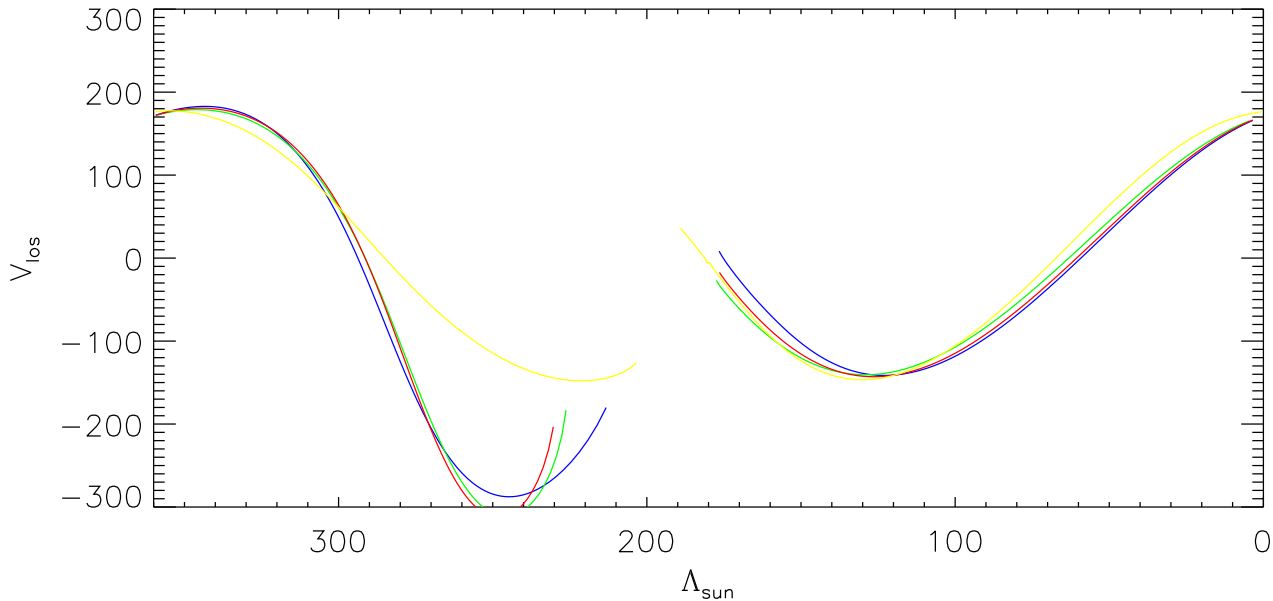}
  \caption[]
  {Line of sight velocity of
the Sagittarius stream as viewed from the sun as a function of its
longitude on the sky. The colours for the models are as previously:
MOND is blue, L05 is green 
and f09CDM is red. The extra line shown is for a {\it prolate} CDM
model (yellow line) with $q=1.25$. It is the yellow line which
provides the best fit to the velocity
data for the Sgr stream, particularly the leading arm data (left most
curves on the plot). Notice that the MOND model (blue line) agrees
well with the L05 and f09CDM models as before (red and green lines),
but that all of the models differ greatly from the prolate model.}
\label{fig:mondvelocity}
\end{center}    
\end{figure*}

\subsection{The Sagittarius leading arm velocity data}\label{sec:velocity}
There has been some debate in the literature over the velocity data
for the Sagittarius dwarf leading arm. We have so far in this paper
referred only to the spatial data from the Sagittarius stream and not
the velocity data.

\citet{2004ApJ...610L..97H} has recently pointed out that the leading
arm velocity data are inconsistent with a halo flattening of $q=0.9$
advocated by \citet{2005ApJ...619..800J} and suggest that a prolate
halo ($q\sim1.25$) would provide a better fit if the velocity data are
taken into account. However, \citet{2005ApJ...619..800J} argue that
a prolate halo leads to an incorrect value for the orbital pole
precession; in fact a prolate halo causes the orbit to precess in the
{\it opposite} direction to that observed. They point out that it is
difficult to obtain the correct 
amount of pole precession without altering the underlying potential
(as can be seen in the pole precession plots in figure
\ref{fig:mondorbits}); whereas one could conceivably alter the
velocities of the stars in the plane of the stream through second
order effects such as dynamical friction
\citep{2005ApJ...619..807L}. Could MOND perhaps reconcile the
discrepant leading arm velocity data? 

Figure \ref{fig:mondvelocity} shows the the line of sight velocity of
the Sagittarius stream as viewed from the sun as a function of its
longitude on the sky. The colours for the models are as previously:
MOND is blue, L05 is green 
and f09CDM is red. The extra line shown is for a {\it prolate} CDM
model (yellow line) with $q=1.25$. It is the yellow line which
provides the best fit to the velocity
data for the Sgr stream, particularly the leading arm data (left most
curves on the plot). Notice that the MOND model (blue line) agrees
well with the L05 and f09CDM models as before (red and green lines),
but that all of the models differ greatly from the prolate model.

MOND does not solve the problem of the discrepant leading arm velocity
data for the Sagittarius stream. As with the spatial data for the
stream, MOND produces a near-identical orbit to the L05 and f09CDM
models. If it can be shown that the Milky Way halo (or any
other galaxy halo) must be prolate, this, as pointed out by
\citet{2004ApJ...610L..97H}, would be difficult to reconcile with MOND.

\section{Discussion}\label{sec:discussion}

\subsection{Model assumptions}
Perhaps the biggest assumption in this work is the highly specific
choice of potential for the galaxy disc. While this could be
problematic for a detailed study of satellite orbits in the Milky Way,
we have implicitly considered the case of maximal precession in MOND.
It would be difficult to imagine a more axisymmetric potential than an
infinitesimally thin disc. This does mean, then, that should a
dark matter halo be discovered which is significantly more oblate than
$q=0.9$, this would be difficult to reconcile with MOND.

We have also neglected dynamical friction and not performed a detailed
N-body simulation of the stripping of stars from the Sagittarius dwarf
galaxy. As such, this work should not be taken as conclusive evidence
that the Sagittarius stream is consistent with MOND. 

Finally, we should note that MOND is probably the best-studied but not
the only alternative gravity theory (see
e.g. \citet{2001PhRvD..63d3503D} and \citet{Moffat:2004ba}). Some of
these theories predict a Keplerian fall off at large radii in the
rotation curve, similar to CDM models (see
e.g. \citet{Moffat:2004ba}). These may be even more difficult to rule
out using tidal streams.

\subsection{Why does such a flat mass distribution in MOND produce
  such a round potential?}

This point has been discussed in some detail in
\citet{2001MNRAS.326.1261M}, but is perhaps best illustrated by direct
integration of equation \ref{eqn:mondequation} for a Kuzmin disc. In
the deep-MOND limit, $\mu(x) \rightarrow x$ and we find from equations
\ref{eqn:mondequation} and \ref{eqn:kuzmin}:

\begin{equation}
\Phi \simeq \frac{(MGa_0)^{1/2}}{2}\ln(R^2+(|z|+a)^2)
\label{eqn:mondpotential}
\end{equation}

Which is very nearly identical to the flattened logarithmic halo (see
equation \ref{eqn:loghalo}). Thus, highly flattened {\it mass
  distributions} in MOND do not produce highly flattened
potentials. In fact, it is the spherical nature of the Kuzmin
potential in the deep-MOND limit which leads to MOND producing
slightly too little precession in the Sagittarius orbit (see figure
\ref{fig:mondorbits}). From figure \ref{fig:rotationcurves}, we can
see that the MOND rotation curve (blue line) is flat beyond $\sim
20$kpc, indicating that it is then in the deep-MOND limit. The other
rotation curves for the CDM models are all falling at these radii
rather than flat - a property which cannot be achieved in MOND, since
it is a theory set up to produce flat rather than falling rotation
curves at large radii. Thus interior to $\sim 20$kpc, all of the
models agree quite well in their rotation curves and the corresponding
orbit for the Sagittarius {\it trailing} arm looks very
similar. This can be seen in the good agreement from -1 to 0Gyrs in
the orbital pole precessions (figure \ref{fig:mondorbits} top right
panel). However, the leading arm orbit - which can be seen in the pole
precession plot from 0 to 1Gyrs - does not agree so well. For this
part of the orbit, the Sagittarius dwarf moves out towards apocentre
and samples the region of the potential where the Milky Way is fully
in the MOND regime and where the rotation curve (in MOND) is flat and
near-spherical.

\subsection{Exploiting flattened elliptical galaxy potentials}

\citet{1994ApJ...427...86B} and \citet{2002ApJ...577..183B} have
recently shown that the observed flattening of hot X-ray gas in
elliptical galaxies may also be used to place tight constraints on
MOND. They argue that, if the hot gas in NGC720
is in hydrostatic equilibrium, then
$\underline{\nabla}p_{gas}=-\rho\underline{\nabla}\Phi$ which implies that
$\underline{\nabla}\rho\times
\underline{\nabla}\Phi=\underline{0}$. Thus, the X-ray isophotes from
the hot gas in NGC720 trace the gas density, which in turn traces the
underlying gravitational potential. By deprojecting the stellar
potential they show that the stars in NGC720 cannot produce a flat enough
potential in MOND to produce the observed X-ray isophotes. 

While they have to assume hydrostatic equilibrium and some simple form
for the deprojected stellar potential, their assumptions are quite
conservative. They find that MOND produces potentials
which are too spherical at large radii - similar to what {\it may} be
the case here for the leading arm Sagittarius dwarf debris.

\subsection{What are the prospects for constraining MOND using tidal
  streams?}

Cold dark matter halos, as modelled in N-body cosmological simulations,
are typically 2:1 triaxial systems. This would make the Milky Way with
$q \sim 0.9$ quite rare. However recent simulations by
\citet{Kazantzidis:2004vu} showed that the dissipation
of baryons changes the inner galactic halos to be nearly
spherical and consistent with the flattening predicted
from the Sagittarius stream. The amount of flattening depends
sensitively on the fraction of baryons that undergoes slow
dissipation to form the galactic disk.

MOND mimics halos with $q \sim 0.9$, while CDM
produces similar halos as a result of gas cooling and galaxy
formation. This will make it difficult to differentiate between MOND
and CDM theories using halo flattening, even with a large statistical
sample of halo shapes.

\section{Conclusions}\label{sec:conclusion}

We have compared orbits in a thin axisymmetric disc potential in MOND to
those in a thin disc plus near-spherical dark matter halo predicted by
$\Lambda$CDM cosmology. We demonstrated that the amount of orbital
precession in MOND is very nearly identical to a similar CDM galaxy
with a logarithmic-halo with flattening q=0.9, consistent with recent
constraints from the Sagittarius stream. Since very flattened mass
distributions in MOND produce more spheroidal potentials than in
standard Newtonian mechanics, we have shown that it will be very difficult
to use the tidal debris from streams to distinguish between a MOND
galaxy and a standard CDM galaxy with a mildly oblate halo. 

If a galaxy can be found with either a prolate halo, or one which is
more oblate than $q \sim 0.9$ this would rule out MOND as a viable
theory. Improved data from the leading arm of the Sagittarius dwarf -
which samples the Galactic potential at large radii - could rule out
MOND if the orbital pole precession can be determined to an accuracy
of the order of $\pm 1^o$.

\section{Acknowledgements}
We would like to thank Mordehai Milgrom, Hongsheng Zhao and Jacob
Bekenstein for useful comments, and the referee, Kathryn Johnston, for
providing us with her best-fit orbit for the Sagittarius dwarf, for
a careful reading of the first draft of this paper and for useful
comments which led to this final version.


\bigskip
\vfil
\bibliographystyle{mn2e}
\bibliography{refs}
 
\end{document}